\documentclass[a4paper,11pt]{article}
\usepackage{graphicx}
\usepackage[dvipdfm]{color}
\usepackage[varg]{txfonts}
\usepackage{indentfirst}
\title{One-loop analysis of the 
four-Fermi contribution to the atomic EDM within RPVMSSM}
\author{Nodoka YAMANAKA, Toru SATO and Takahiro KUBOTA
\\
{\small  {\it Department of Physics, Osaka University, Osaka 560-0043, Japan} }}
\date{\empty}
\begin{document}
\maketitle

\def \del{\partial}

\begin{abstract}
The contribution in the R-parity violating (RPV) Minimal Supersymmetric Standard Model (MSSM) to the electric dipole moment (EDM) of $^{199}$Hg at the one-loop level is 
 evaluated. At the one-loop level, the $^{199}$Hg EDM 
 receives  RPV contribution whose couplings are of a  
different type from
the tree level analysis. This contribution 
 is shown to 
be constrained by using the limit to the CP-odd electron-nucleon (e-N) interaction given by the recent result of $^{199}$Hg EDM experiment.
\end{abstract}

\section{Introduction}
Among the New Physics candidates, the supersymmetric extension of the  Standard Model (SM), the MSSM is the  most leading one  for several reasons. The MSSM has many advantages, namely the cancellation of 
 quadratic divergences in radiative corrections due to the Higgs field, the radiative breaking of the electroweak symmetry, possibility of explaining the dark matter 
with stable lightest sparticle and so forth.

 In MSSM, the lepton  and baryon numbers are not necessarily conserved and  the conservation of {\it R-parity} is  assumed  to prohibit fast nucleon decays.  However, 
 such an  assumption is {\it ad hoc}, and we must investigate R-parity violation 
 with a close look at  phenomenological constraints.    We should also note that the coupling constants of RPV interactions are in general complex and therefore provide us with new sources of CP-violation besides the Kobayashi-Maskawa phase in SM. As is known rather well, the baryon/photon ratio in our Universe is hardly explained within SM.  Those phases  of  RPV couplings are of great interest in this respect and should be scrutinized with the help of CP violation observables.

 Among  observables sensitive to CP violation, 
 electric dipole moment (EDM) may help us reveal mechanism of CP violation.  
 Many EDM measurements  including those of neutron,  electron and atoms, have been 
performed until now. In  SM, the EDM is predicted to be so 
small  that it turns out 
to be a very efficient probe of New Physics. In this 
 talk we focus on the atomic EDM 
which comes from P- and CP-violating e-N interactions 
\cite{flambaum}.

Herczeg \cite{herczeg} pointed out that  P- and CP-violating e-N  interactions could be produced by sneutrino exchange between  electron and  quark via 
RPV-interactions {\it at  the tree level}.  He evaluated such an exchange effect and determined general form of e-N  interaction. By comparing it with the  atomic EDM ($^{205}$Tl) data then available, Herczeg obtained constraints of the RPV couplings.

Recently,  the EDM of $^{199}$Hg 
was measured by the group of Seattle  \cite{griffith}, and  the new upper limit was translated into upper bounds on fundamental CP violating parameters. In particular, those of P- and CP-odd e-N interactions have been constrained with unparalleled precision.
The purpose  of the present work is 
 to confront the P- and CP-odd  e-N 
interactions given in \cite{griffith}
 with the theoretical computation  
{\it at the one-loop level}  in RPVMSSM.
 We would like to point out that our one-loop analysis gives constraints 
on a combination of RPV couplings which  are different from Herczeg's.
Our discussion is organized as follows. We first briefly review the R-parity violation and calculate its contribution to the atomic EDM at the one-loop level, then compare it with the  new experimental data \cite{griffith} to obtain bounds on RPV interactions, and finally summarize  our results.

\section{P- and  CP-odd e-N interactions within RPVMSSM}

 We will consider the P- and CP-violating interactions in RPVMSSM whose interaction Lagrangian is given by
\begin{eqnarray}
{\cal L}_{R\hskip-.2cm /}&=&
-\frac{1}{2}\sum _{ijk}\lambda _{ijk}\{
\tilde e^{\dag}_{Rk}\bar \nu ^{c}_{i}P_{L}e_{j}+\tilde e_{Lj}\bar e_{k}P_{L}
\nu _{i}+
\tilde \nu _{i}\bar e_{k}P_{L}e_{j}
\nonumber \\
& & -\tilde e^{\dag}_{Rk}\bar e^{c}_{i}P_{L}\nu _{j}
-\tilde e_{Li}\bar e _{k}P_{L}\nu _{j}-\tilde \nu _{j}\bar e_{k}P_{L}e_{i}
\}
\nonumber \\
& & -\sum _{ij  k}\lambda '_{ijk}\{
\tilde d_{Rk}^{\dag}\bar \nu _{i}^{c}P_{L}d_{j}+\tilde d_{Lj}\bar d_{k}P_{L}\nu _{i}+\tilde \nu _{i}\bar d_{k}P_{L}d_{j}
\nonumber \\
& & -\tilde d^{\dag}_{Rk}\bar e_{i}^{c}P_{L}u_{j}-\tilde e_{Li}\bar d_{k}P_{L}u_{j}-\tilde u_{Lj}\bar d_{k}P_{L}e_{i}\}
+{\rm h. c.}
\label{eq:rpvinteraction}
\end{eqnarray}
with $P_{L}=(1-\gamma ^{5})/2$ and $\lambda _{ijk}=-\lambda _{jik}$.
 Here the indices $i,j,k =1,2,3$ indicate the generations. 
Many of these RPV interaction terms are constrained phenomenologically \cite{chemtob}.

The general form of the P- and CP-odd 
e-N   interaction contributing to the $^{199}$Hg EDM is given by \cite{flambaum}
\begin{equation}
H_{P\hspace{-.4em}/ \, ,T\hspace{-.4em}/ } = \frac{G_F}{\sqrt{2}}  \sum_{N=p,n}
\left\{
C_N^{SP} \bar N N \bar e i\gamma_5 e + C_N^{PS} \bar N i\gamma_5 N \bar e e + C_N^T \epsilon_{\mu \nu \rho \sigma } \bar N \sigma^{\mu \nu} N \bar e \sigma^{\rho \sigma} e
\right\} \ ,
\label{eq:enint}
\end{equation}
where $C_N^{SP}$, $C_N^{PS}$  and  $C_N^T$ are real parameters. 
 Let us recall that Herczeg \cite{herczeg} evaluated the sneutrino exchange tree-diagram, studied the contributions to $C^{SP}_{N}$ and $C_{N}^{PS}$ and deduced phenomenological constraints on 
\begin{eqnarray}
\sum _{k=1,2,3} \sum _{j=2,3}
 {\rm Im}\left ( \lambda ^{*}_{1j1}\lambda '_{jkk}\right ).
 \label{eq:tree}
\end{eqnarray}

From the recent experimental data of $^{199}$Hg   EDM ($d_{\rm Hg} = (0.49 \pm 1.29 \pm 0.76 ) \times 10^{-29} e\, cm$ \cite{griffith}), we obtain the following bounds on P- and CP-odd 
e-N  interactions \cite{flambaum,griffith}:
\begin{eqnarray}
\left| \, 0.40 C_p^{SP} + 0.60 C_n^{SP} \right| &<& 5.2 \times 10^{-8} \ ,
\label{eq:limcsp}\\
\left| \, 0.24 C_p^{PS} + 0.76 C_n^{PS} \right| &<& 5.1 \times 10^{-7} \ ,
\label{eq:limcps}
\\
\left| \, 0.24 C_p^T + 0.76 C_n^T \right| &<& 1.5 \times 10^{-9} \ .
\label{eq:limct}
\end{eqnarray}
See Ref. \cite{flambaum} for details of derivation of (\ref{eq:limcsp}) $-$ (\ref{eq:limct}).
The SM contributes to these parameters at the level of $10^{-16}$ \cite{he}, but the one-loop contribution of some RPV interactions (not constrained at the tree level) can be much larger than these limits.

\begin{figure}[h]
\begin{center}
\includegraphics[height=48mm]{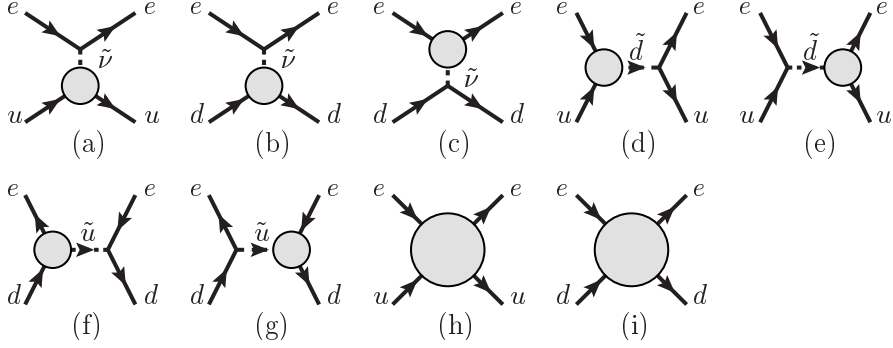}
\caption{\label{fig:loop_class} Possible one-loop corrections to P- and CP-odd e-N interactions within RPVMSSM.}
\end{center}
\end{figure}

\begin{figure}[t]
\begin{center}
\includegraphics[height=32mm]{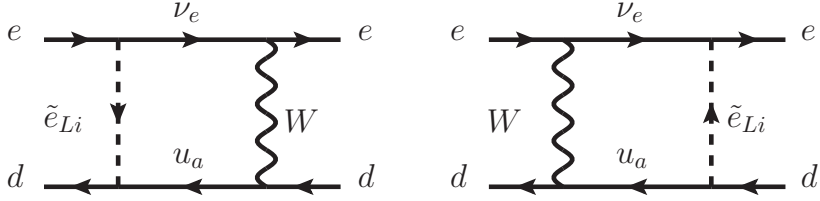}
\caption{\label{fig:box} Box diagrams contributing to P- and CP-odd e-N interactions within RPVMSSM.}
\end{center}
\end{figure}

With the RPV interactions (\ref{eq:rpvinteraction}), we can construct one-loop corrections to the P- and CP-odd 
e-N  interactions as shown in Fig.\ref{fig:loop_class}.
The vertex corrections  (a) $\sim$ (g) do not have contributions to the EDM   for 
the following reasons: vertex correction can be absorbed in the renormalization of RPV interactions used at the tree level; the RPV couplings give no imaginary part;   suppression due to Yukawa couplings. Of course the contributions of one-loop graphs with RPV couplings (\ref{eq:tree}) are already strongly constrained from tree level analysis \cite{herczeg}, so we don't consider them. 
After {\rm detailed} analysis, we see that there 
 are only two significant box diagrams, which are shown in Fig.\ref{fig:box}. These two contributions are  hermitian conjugate of each other and they are summed up into
\begin{eqnarray}
{\cal M}&=&-\frac{e^{2}}{2{\rm sin}^{2}\theta _{W}}\sum _{i}\sum _{a}
\lambda _{1i1}^{*}\lambda _{ia1}'\: V_{a1}\:I_{ai}\left [
\bar e i \gamma ^{5}e \cdot \bar d d -\bar e e \cdot \bar d i \gamma ^{5}d 
+(\mbox{P-even terms})
\right ]
\nonumber \\
& & +\mbox{h.c.}
\label{eq:amplitude}
\end{eqnarray}
where $i$ and $a$ are flavor indices.  $I_{ai}$ is the loop integral of the box diagram
\begin{eqnarray}
I_{ai}=
\frac{1}{4(4\pi )^{2}}\left \{
f(m_{\tilde e_{Li}}, m_{W}, m_{u_{a}})+
f(m_{W}, m_{\tilde e_{Li}}, m_{u_{a}})+f(m_{u_{a}}, m_{W}, m_{\tilde e_{Li}})
\right \},
\label{eq:iai}
\end{eqnarray}
where
$f(a, b, c) \equiv  a^{2}{\log }a^{2}/(a^{2}-b^{2})(a^{2}-c^{2})$. 
We shall discuss the dependence of (\ref{eq:iai}) on s-particle mass later.
From (\ref{eq:limcsp}) and (\ref{eq:limcps}) we can see  that
$C^{SP}_{N}$ is more severely constrained than $C^{PS}_{N}$ and so we focus on the first term in (\ref{eq:amplitude}). By using nucleon matrix elements 
we can derive from (\ref{eq:amplitude}) the  coefficients $C^{SP}_{N}$ on the nucleon level Hamiltonian (\ref{eq:enint}) as follows:
\begin{eqnarray}
C^{SP}_{p}&=&8m_{W}^{2}\sum _{i}\sum _{a}{\rm Im}(
\lambda _{1i1}^{*}\lambda '_{ia1})V_{a1}\:I_{ai}\langle p \vert \bar d d \vert p \rangle ,
\label{eq:cspp}
\\
C^{SP}_{n}&=&
8m_{W}^{2}\sum _{i}\sum _{a}{\rm Im}(
\lambda _{1i1}^{*}\lambda '_{ia1})V_{a1}\:I_{ai}\langle p \vert \bar u u  \vert p \rangle , 
\label{eq:cspn}
\end{eqnarray}
where   
 $\langle p \vert \bar u u \vert p \rangle =3.5$ and $\langle p \vert \bar d d \vert p \rangle =2.8$ were calculated by using isospin symmetry \cite{zhitnitsky} and  current quark masses \cite{leutwyler}. It should be noted that the 
 RPV couplings ${\rm Im}(\lambda _{1i1}^{*}\lambda '_{ia1})$ in 
 (\ref{eq:cspp}) and (\ref{eq:cspn}) differ from  those in (\ref{eq:tree}). 
If we could assume that the couplings 
$\sum _{i, a}{\rm Im}(\lambda _{1i1}^{*}\lambda '_{ia1})$ be  much larger than 
(\ref{eq:tree}), then we could  obtain upper limits on RPV interactions 
$\left|  \sum _{i, a}{\rm Im} (\lambda^*_{1i1} \lambda'_{ia1}) \right|$  
by putting (\ref{eq:cspp}) and (\ref{eq:cspn}) into (\ref{eq:limcsp}). 

\begin{figure}[h]
\begin{tabular}{cc}
\begin{minipage}{0.5\hsize}
\begin{center}
\includegraphics[height=43mm]{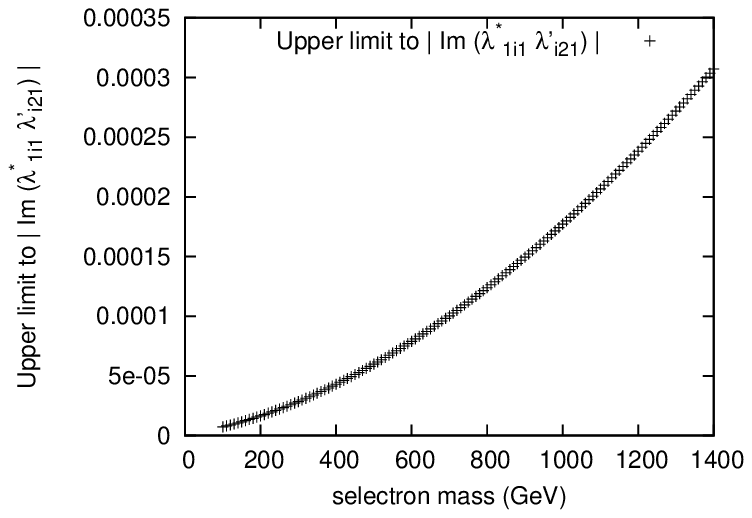}
\label{fig:winter}
\end{center}
\end{minipage}
\begin{minipage}{0.5\hsize}
\begin{center}
\includegraphics[height=43mm]{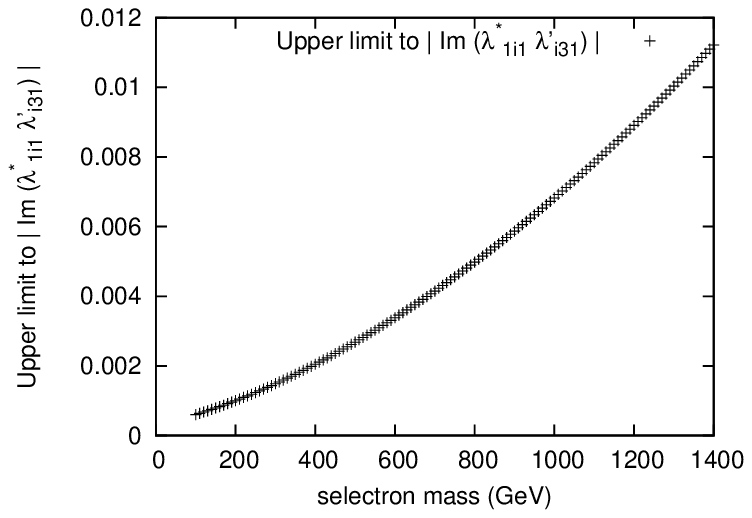}
\label{fig:fall}
\end{center}
\end{minipage}
\end{tabular}
\caption{
\label{fig:limdepdce}
The region under the solid line is allowed. 
In the left (right ) figure, the 
contribution from the charm (top) quark in the loop 
is taken as the most dominant.
}
\end{figure}

The allowed regions of 
$\left| {\rm Im} (\lambda^*_{1i1} \lambda'_{ia1}) \right|$
from the constraint  (\ref{eq:limcsp}) are shown in Fig. \ref{fig:limdepdce}.

\vskip2mm
\vskip2mm
\noindent
In drawing Fig. \ref{fig:limdepdce} we assume implicitly that only one term 
in the summation in  (\ref{eq:cspp}) and (\ref{eq:cspn}) 
is much dominant over the others. In the left (right ) figure in 
Fig. \ref{fig:limdepdce}, the loop diagram in which the charm (top) quark 
is propagating is taken as the most dominant.
As we can see, the upper limits become looser as the selectron mass 
($m_{\tilde e_{Li}}$) grows. By setting SUSY particle masses equal to 
100 GeV, we obtain the limits to the RPV couplings as shown in Table 
\ref{rpvlimit}. We show also the current limits to the RPV couplings 
from other experimental data \cite{chemtob}.

\begin{table}[h]
\begin{center}
\begin{tabular}{lcc}
\hline
RPV couplings & Limit from $^{199}$Hg EDM & Limits in Ref.  \cite{chemtob} \\
\hline
$\left| {\rm Im} (\lambda^*_{121} \lambda'_{221}) \right|$ & $7.3 \times 10^{-6}$ & $ 4.8 \times 10^{-4}$ \\
$\left| {\rm Im} (\lambda^*_{131} \lambda'_{321}) \right|$ & $7.3 \times 10^{-6}$ & $6.0 \times 10^{-4}$ \\
$\left| {\rm Im} (\lambda^*_{121} \lambda'_{231}) \right|$ & $6.0 \times 10^{-4}$ & $8.8 \times 10^{-3}$ \\
$\left| {\rm Im} (\lambda^*_{131} \lambda'_{331}) \right|$ & $6.0 \times 10^{-4}$ & $6.0 \times 10^{-3}$ \\
\hline
\end{tabular}
\end{center}
\caption{Comparison 
of the  upper bounds on RPV couplings from $^{199}$Hg EDM
 with other 
experimental data \cite{chemtob} . 
Sparticle masses 
are assumed as $m _{\tilde e_{Li}}= 100$ GeV.
 }\label{rpvlimit}
\end{table}

The combinations of the RPV couplings constrained in this discussion 
differ from those of the tree level analysis  by Herczeg \cite{herczeg}. 
This is simply because we considered $W$  boson exchange diagrams and  the 
CKM flavor change  are taken into account. 
By comparing our limits with the current phenomenological bounds on the 
RPV interactions from other experiments \cite{chemtob}, we see that 
our one-loop analysis gives  tighter 
constraints by one or two orders of magnitude on the imaginary 
parts of RPV couplings.

\section{Summary}
To summarize our calculation, we have analyzed the RPV contribution of 
the P- and CP-odd e-N interactions at the one-loop level and found 
from the recently updated $^{199}$Hg EDM data  \cite{griffith}   
new  limits on 
$ {\rm Im} (\lambda^*_{121} \lambda'_{221}) $ , 
$ {\rm Im} (\lambda^*_{131} \lambda'_{321}) $ , 
$ {\rm Im} (\lambda^*_{121} \lambda'_{231}) $ , 
and $ {\rm Im} (\lambda^*_{131} \lambda'_{331}) $, 
as shown in Table \ref{rpvlimit}.

\end{document}